\documentclass{aastex}          
\usepackage{spr-astr-addons}    

\begin{document}
\title{UVMag: stellar formation, evolution, structure and environment with space UV and visible spectropolarimetry}
\shorttitle{UVMag}
\shortauthors{Neiner et al.}

\author{C. Neiner} 
\affil{LESIA, UMR 8109 du CNRS, Observatoire de Paris, UPMC, Univ. Paris Diderot, 5 place Jules Janssen, 92195 Meudon Cedex, France}
\affil{coralie.neiner@obspm.fr} 
\author{D. Baade}
\affil{ESO, Karl-Schwarzschild-Str. 2, D-85748 Garching, Germany}
\author{A. Fullerton}
\affil{Space Telescope Science Institute, Baltimore, MD, USA}
\author{C. Gry}
\affil{Laboratoire d'Astrophysique de Marseille, Universit\'e d'Aix-Marseille \& CNRS, 38 rue Fr\'ed\'eric Joliot-Curie, 13388, Marseille Cedex 13, France}
\author{G. Hussain}
\affil{ESO, Karl-Schwarzschild-Str. 2, D-85748 Garching, Germany}
\author{A. L\`ebre}
\author{J. Morin}
\affil{LUPM - UMR 5299 - Universit\'e Montpellier II/CNRS, 34095, Montpellier, France}
\author{P. Petit}
\affil{CNRS, IRAP, 14 Avenue Edouard Belin, 31400, Toulouse, France; Universit\'e de Toulouse, UPS-OMP, IRAP, Toulouse, France}
\author{J. O. Sundqvist}
\affil{Universit\"atssternwarte M\"unchen, Scheinerstr. 1, D-81679 M\"unchen, Germany}
\author{A. ud-Doula}
\affil{Penn State Worthington Scranton, Dunmore, PA, 18512, USA}
\author{A.~A. Vidotto}
\affil{SUPA, School of Physics and Astronomy, University of St Andrews, St Andrews, Scotland KY16 9SS, UK}
\author{G.~A. Wade}
\affil{Department of Physics, Royal Military College of Canada, PO Box 17000, Stn Forces Kingston, Ontario K7K 7B4 Canada}
\and
\author{the UVMag consortium} 


\begin{abstract}
Important insights into the formation, structure, evolution and environment of
all types of stars can be obtained through the measurement of their winds and 
possible magnetospheres. However, this has hardly been done up to now mainly
because of the lack of UV instrumentation available for long periods of time. To
reach this aim, we have designed UVMag, an M-size space mission equipped with a
high-resolution spectropolarimeter working in the UV and visible spectral range.
The UV domain is crucial in stellar physics as it is very rich in atomic and
molecular lines and contains most of the flux of hot stars. Moreover, covering
the UV and visible spectral domains at the same time will allow us to study the
star and its environment simultaneously. Adding polarimetric power to the
spectrograph will multiply tenfold the capabilities of extracting information on
stellar magnetospheres, winds, disks, and magnetic fields. Examples of science
objectives that can be reached with UVMag are presented for pre-main sequence,
main sequence and evolved stars. They will cast new light  onto stellar physics
by addressing many exciting and important questions. UVMag is currently
undergoing a Research \& Technology study and will be proposed at the
forthcoming ESA call for M-size missions. This spectropolarimeter could also be
installed on a large UV and visible observatory (e.g. NASA's LUVOIR project)
within a suite of instruments.
\end{abstract}


\section{Introduction}

During the formation and the entire life of stars, several physical processes
influence their dynamics and thus impact their evolution. In particular magnetic
fields and winds are key factors. They directly affect the internal structure of
massive stars \citep{maeder2009} and cool stars \citep{chabrier2007}, the
transport of angular momentum \citep{mathis2013}, the formation and mass
accretion phases \citep{bouvier2007}, and of course the direct circumstellar
environment. They thus drive stellar evolution. Therefore the aim of the UVMag
(UltraViolet and Magnetism) space project is to study the formation, structure,
evolution and environment of all types of stars in particular through the
measurement of their winds and possible magnetospheres, i.e. through the
association of spectropolarimetry and spectroscopy in the UV and visible
domains.

The UV domain is crucial in stellar physics because it is particularly rich in
atomic and molecular transitions, especially resonance lines, and covers the
region in which the intrinsic spectral energy distribution of hot stars
peaks. Resonance lines are excellent diagnostics of the physical state and
chemical composition of low-density plasmas. They are less likely to depopulate
in low density environments such as chromospheres, circumstellar shells, stellar
winds, nebulae and the interstellar medium, and so remain the only useful
diagnostics in most of these environments. Another advantage of observing in the
UV is the extreme sensitivity of the Planck function to the presence of small
amounts of hot gas in dominantly cool environments. This allows the detection
and monitoring of various phenomena that would otherwise be difficult to
observe: accretion continua in young stars, magnetic activity, chromospheric
heating, coronae, active regions on cool stars, and intrinsically faint, but
hot, companions of cool stars. The UV domain is also exquisite for many
other aspects. For example it is the domain in which Sun-like stars exhibit
their hostility (or not) to Earth-like life, population III stars must
have shone the brightest, accretion processes convert much kinetic energy into
radiation which strongly impacts stellar formation and evolution, the ``Fe
curtain'' features respond to changes in local irradiation, etc. Moreover, many
light scattering and polarising processes are stronger at UV wavelengths.  

In addition, most cool stars and some fraction of the hot stars are magnetic
\citep{petit2014} and their magnetic field interacts with their wind and
environment, modifies their structure and surface abundances, produces temperature variations, and contributes to
the internal and external transport of angular momentum. Magnetic fields can be
measured via the polarisation of the stellar light \citep[see
e.g.][]{landstreet2009}. With spectropolarimetry, one can thus address with
unprecedented detail these important issues in stellar physics, from stellar
magnetic fields to surface inhomogeneities, surface differential rotation to
activity cycles and magnetic braking, from microscopic diffusion to turbulence,
convection and circulation in stellar interiors, from abundances and pulsations
in stellar atmospheres to stellar winds and accretion disks, from the early
phases of stellar formation to the late stages of stellar evolution, from
extended circumstellar environments to distant interstellar medium. Moreover,
measuring polarisation directly in the UV wind-sensitive lines has never been
done.  The combination of UV polarisation and velocity information from line
profiles, which can be used as a proxy to trace the wind density, would permit a
3 dimensional tomographic view of the magnetic field lines to be constructed.
Finally, polarimetry is not restricted to magnetic fields only. The scope of
stellar polarimetry is much broader, in particular with respect to linear
polarisation and depolarisation processes in circumstellar environments, e.g. in
accretion or decretion disks, or from exoplanets.

The UVMag project presented here has been designed to address all these exciting
and important questions. For this, we need to obtain simultaneous UV and visible
spectropolarimetry, continuously over at least one stellar rotation period. Of
course, the UV domain requires a space mission. Therefore the UVMag consortium
proposes to build a dedicated M-size space mission with a telescope of 1.3 m and
UV and visible spectropolarimetric capabilities. The spectropolarimetric
capability of UVMag, both in the UV and visible wavelength domains, will nicely
complement its spectrograph to multiply tenfold the capabilities of extracting
information on magnetospheres (i.e. circumstellar material retained within the
magnetic field lines), winds, disks, and magnetic fields. Moreover, the better
sampling afforded by a space-based instrument will allow us to obtain nearly
continuous time-series of short-cadence measurements. Such time-series of data document
phenomena on stars that can be impulsive (flares, infall),  periodic
(pulsations, migration of spots, corotating clouds), quasi-periodic (evolution
of clumps and shock-induced inhomogeneities from hot winds), or gradual
(evolution of spots). 

\section{Science objectives}

A UV and visible spectropolarimeter will provide a very powerful tool to study
most aspects of stellar physics in general and in particular for stellar
formation, structure and evolution as well as for the stellar environment. The
full science case is available at http://lesia.obspm.fr/UVMag. A non-exhaustive
list of examples is provided below, assuming an M-size space mission
dedicated to stellar physics. If the spectropolarimeter were to be installed on
a larger telescope, its scientific scope would be extended to many more topics
such as white dwarfs.

\subsection{Pre-main sequence stars}

\subsubsection{T Tauri stars}

T Tauri stars (TTS) are young, low to intermediate mass (0.5-2.5 M$_\odot$)
pre-main-sequence (PMS) G to M stars, many of which are still surrounded by
substantial disks. PMS models show that TTS interiors are either fully
convective or have an outer convective envelope, depending on the stellar age
and mass. They develop a radiative core as they approach the main sequence. TTS
with masses greater than 1.2 M$_\odot$ then become fully radiative at the age of
20-30 Myr. The exact timescale on which this happens depends on the stellar mass
and the PMS track used
\citep[e.g.][]{palla1993,chabrier1997,siess2000,behrend2001}.

It is necessary to better understand the role of the convective phase that stars
experience before reaching the birth-line on the fossil field. A first
observational result is that many intermediate-mass TTS (IMTTS) are indeed
magnetic \citep{hussain2009}. However, magnetic observations of this type of
stars have only just started \citep{hussain2014}.

TTS show variability at both short and long timescales, and over a range of
wavelengths, from radio to X-ray. Many systems display excess UV continuum and
Balmer emission lines, indicating that accretion from large  circumstellar disks
is ongoing. Stars that are accreting are called classical TTS (cTTS). As these
systems evolve onto the main sequence, the gas and dust in their disks dissipate
and can condense into planetary systems. ``Naked'' or weak-lined TTS (wTTS)
represent a later evolutionary stage, have more evolved disks and do no longer
accrete \citep{rebull2006,cieza2007}. Determining the timescales over which they
stop accreting material from their surrounding disks is a key area in TTS
research \citep[e.g.][]{natta2007,olofsson2009,oliveira2011}.

\cite{brown1981} were the first to reveal that TTS host magnetic fields. These
fields have later been shown to play a critical role in explaining key
properties of TTS through magnetospheric accretion
\citep{camenzind1990,konigl1991}. For a comprehensive review, see
\cite{hussain2012}. In brief, the stellar magnetic field  truncates the inner
edge of the accretion disk either at or within the Keplerian corotation radius.
Disk material is then channelled along these magnetic field lines onto the
central star causing a shock above the stellar surface before the material can
settle onto the star. Several analytical models of magnetospheric accretion have
been developed, assuming a dipolar stellar field
\citep{konigl1991,shu1994,choi1996}. More recent simulations of accretion in TTS
account for the complex field topologies that have been discovered by
spectropolarimetric studies \citep[e.g.][]{romanova2011}.

A long-standing puzzle has been the reason for slow rotation in TTS (P$_{\rm
rot}$=7-10 days). As TTS contract and accrete material they would be expected to
spin up and rotate at a few percent of their breakup speed. Instead the magnetic
interaction between the star and the slowly rotating disk is responsible for
slowing down the star \citep{Ghosh1978,shu1994}. The efficiency of this braking
process depends on the size and geometry of the magnetic field.

Since they are accreting, cTTS should host very strong accretion-powered winds
and can thus also lose angular momentum through their winds
\citep[e.g.][]{hartmann1982,matt2005,matt2008}. The clumpy character of
magnetospheric accretion can provide the required turbulent energy to the stars
to drive their winds with mass-loss rates 10000 times weaker than the accretion
rates \citep[e.g.][]{cranmer2008,cranmer2009}. However, the latest calculations
suggest that winds alone may not be sufficient to explain the observed rotation
rates, with a combination of both disk locking and accretion-powered winds
required to explain the observed angular momentum properties of these systems
\citep{matt2010,zanni2011,brickhouse2012}.

Magnetospheric models predict that the longer a star is coupled to its disk, the
slower it should rotate, a prediction supported by Spitzer observations of young
stellar clusters \citep{cieza2007}. Once accretion stops the star is then
presumably free to spin up. The timescales for disk locking are therefore very
important for explaining the distribution of rotation periods on the zero age
main sequence \citep{bouvier1997}.

TTS magnetic fields are also important for the evolution of the entire
system. Strong X-ray and UV heating from coronal heating, large energetic flares
and coronal mass ejections can be responsible for driving the evolution of disks
through irradiation and photo-evaporation
\citep[e.g.][]{gorti2009,ercolano2009,owen2011}. Furthermore, the strength and
geometry of the stellar magnetic field determines how far the inner disk can
extend and therefore sets a limit to planet migration
\citep{romanova2006,long2011}.

To explain many of the above processes we need to understand TTS magnetic fields
- the underlying driver of magnetospheric accretion. By probing the strength and
geometry of magnetic fields in TTS, we can start to understand how these
properties depend on stellar parameters \citep[see][]{gregory2012}  and to
develop more realistic  models of magnetospheric accretion and winds
\citep[e.g.][]{romanova2011}. This is essential to develop a better
understanding of stellar and planetary formation.

In weakly accreting TTS systems that are in the process of ``switching off''
accretion, accretion is best detected with UV diagnostics
\cite[e.g.][]{ingleby2011}. Combined spectro-polarimetric and UV studies of
selected well studied systems would be important to evaluate the role of the
stellar magnetic field on ``switching off'' accretion in systems transitioning
into naked or weak TTS. 

In particular, UV observations possess a variety of diagnostics that can be used
to derive key insights into the accretion state of TTS: from Si{\sc iv}, C{\sc
iv}, N{\sc v} emission that points to hot gas from accretion shocks and
absorption from outflows, to H$_2$ and CO emission originating in disks (at
temperatures of 300-600 K and $\sim$2500 K respectively). Moderate resolution
spectra are necessary to resolve the structure in the lines forming in the
accretion shocks and outflows and test magnetospheric accretion theories.
Contemporaneous spectropolarimetric visible and UV studies would enable more
detailed, quantitative, tests of the magnetospheric accretion models than have
been possible to date. At the same time the spectra would allow us to check if
the studied TTS are binaries, which is an important aspect for the
interpretation of many observations.

Recently \cite{gomezdecastro2012} determined UV- and X-ray-normalised fluxes to
study the extent and properties of the TTS magnetospheres as a class. They found
that the normalised fluxes are correlated in a different way than those of the
main sequence cool stars. In particular there is a very significant excess
emission in O{\sc i} in the TTS caused by recombination radiation from the disk
atmosphere after photoionization by extreme UV radiation, and the stronger the
X-ray surface flux is, the weaker the observed UV flux. This last behaviour is
counter-intuitive within the framework of stellar dynamo theory and suggests
that UV emission can be produced in the extended and dense stellar magnetosphere
directly driven by local collisional processes. The brown dwarf 2MASS
J12073346-3332539 has been found to follow the same flux-flux relations as the
TTS. Thus, TTS-normalised flux scaling laws seem to be extendable to the brown
dwarf limit and can be used for identification/diagnosis purposes. 

\subsubsection{Herbig Ae/Be stars}

Herbig Ae/Be (HAeBe) stars are the PMS progenitors of A/B stars with masses between 2.5 and 13 M$_\odot$. They are the
higher-mass counterparts of TTS. HAeBe stars are contracting towards the main
sequence and are surrounded by dust and gas left over from their original
molecular cloud. To understand the formation of massive stars, as well as their
magnetic and rotation properties, it is necessary to understand the structure of
the matter around HAeBe stars and its interaction with the star. 

While many pieces of evidence of magnetospheric accretion exist for TTS, it is
not clear if accretion exists and how it operates onto HAeBe stars. Contrary to
TTS, there is no clear indication of magnetospheric accretion, such as veiling,
or the presence of blue-shifted forbidden emission lines (e.g. [O{\sc i}] 6300
\AA). However, HAeBe stars can show UV Balmer excess emission and H$\alpha$
profiles, which appear analogous to those found in accreting TTS
\citep{mendigutia2011}. Inverse P-Cygni profiles and red-shifted circumstellar
absorption features are also observed, mostly in Herbig Ae stars but also some
Herbig Be stars \citep[e.g.][]{boley2009}. However, these features could be
explained in terms of clumpy accretion rather than magnetospheric accretion
\citep[e.g.][]{mora2004}. In addition, \cite{finkenzeller1984} have detected P
Cygni profiles in the H$\alpha$ and Mg{\sc ii} h \& k lines in a large number of
HAeBe stars, which indicate that they possess a stellar wind. The presence of
forbidden emission lines, such as [O{\sc i}] (6300 \AA), also points toward the
existence of a stellar wind \citep{boehm1994}. Moreover, radio emission detected
in HAeBe stars \citep{skinner1993} is predominantly thermal, and in many cases
wind-related. Finally, spectro-interferometric observations, around the
Br$\gamma$ line, of few HAeBe stars are better interpreted with the presence of
an optically thick disk and a stellar wind whose apparent extent is much larger
than the disk extension \citep[e.g.][]{malbet2007}.

However, the origin of the winds and jets associated with HAeBe stars is poorly
understood. \cite{corcoran1997} propose the theory of accretion driven winds.
They argue that the positive correlations observed between the forbidden
emission lines [O{\sc i}] and H$\alpha$, or the IR excess, imply a strong link
between outflows and disk. However, in many HAeBe stars, the absence of
high-mass disks (especially in HBe stars), as well as the evidence of low-mass
accretion rates \citep[e.g.][]{garcialopez2006}, are rather an indication of
passive disks. In HBe stars with their higher radiative flux, especially in the
UV, theories like radiation driven winds might be more appropriate
\citep[e.g.][]{babel1997}.

According to the fossil field hypothesis, a small fraction of the HAeBe stars 
should host strong magnetic fields of simple configuration. In order to test
this proposal, \cite{alecian2012a,alecian2012b} performed a high-resolution
spectropolarimetric survey of 70 HAeBe stars located in the field of the Galaxy.
They found 9 magnetic stars \citep{alecian2012a}, implying an incidence of about
10\%. Four of them have been studied in details, and they find that the fields
are mainly dipolar, strong (from 300 G to 2.1 kG), and stable over more than 5
years \citep[e.g.][]{folsom2008,alecian2009}. These results confirm that a
fossil link exists between the pre-main sequence and main-sequence magnetic
fields in the intermediate-mass stars, and that their magnetic fields must have
been shaped during the star formation. However, magnetic studies of HAeBe stars,
and in particular the roles of fields in mediating accretion, are still in their
infancy. 

Using UV and visible spectropolarimetry would allow us to study the stellar
magnetic field and wind of HAeBe stars to test, e.g., the origin of their wind,
whether their disk is passive or linked to outflows, whether magnetospheric
accretion exists, etc. In particular the variability of the wind and the disk in
HBe stars would probably enable time series of UV spectra to assess the validity
of the hypothesis of a disk origin of the winds.  

\subsection{Hot stars}

Hot (OB) stars dominate the ecology of the universe as cosmic engines via
their extreme output of radiation and matter, not only as supernovae but also
during their entire lifetime with far-reaching consequences. They usually
display strong variability on various time scales due to such phenomena as mass
outflows, rapid rotation, pulsations, magnetism, binarity, radiative
instabilities, and the influence of their circumstellar environment. In
particular this applies to classical and Herbig Be, Bp, $\beta$ Cep, Slowly
Pulsating B (SPB), B[e] and O stars, as well as massive binaries such as the Be
X-ray binaries and those that harbour O-type subdwarf companions, and virtually
all evolved OB stars.

Research in the domain of OB stars has been progressing very rapidly in the last
decade. However, UV spectrographs are missing to study the wind and
magnetospheres of these objects and the current studies \citep[e.g.][]{petit2013}
rely mainly on the IUE archives which contains data for a rather limited number
of stars. Archival IUE data rarely have contemporaneous observations at other
wavelengths while the variability is often non-periodic, and the quality of data
from contemporary facilities operating at non-UV wavelengths is also vastly
improved. Efficient high-resolution ground-based spectropolarimeters (Narval,
ESPaDOnS, HARPSpol) provide important clues about magnetic fields and the
confinement of the circumstellar environment. These instruments, however, can
only supply high S/N measurements in the visible domain of relatively bright stars and can hardly reach
cluster stars.

Spectropolarimetry of fainter and numerous cluster stars could be reached with
observations from space by adding signal from the UV domain, where OB stars emit
most of their light. More importantly, simultaneous UV spectroscopy and visible
spectropolarimetry would provide clear diagnostics for the study of
magnetospheres and circumstellar environments. 

The magnetic fields of high-mass stars are different from those of low-mass
stars \citep[e.g.][]{neiner2007}. They are detected in only 7\% of high-mass
stars \citep{wade2013} and they are structurally much simpler, and internally
much stronger, than the fields of cool stars. In addition, their characteristics
show no clear correlation with basic stellar properties such as age, mass or
rotation \citep[e.g.][]{mathys1997,kochukhov2006,landstreet2007}.  These
characteristics reflect a different field origin: they are fossil fields, i.e.
remnants of field accumulated or generated during star formation, rather than
fields generated by dynamos  \citep[e.g.][]{mestel2001,moss2001,ferrario2006}.
This fossil origin allows us to study how magnetic fields are modified as well
as how they influence the evolution of stellar properties. For example, recent spectropolarimetric observations show that magnetic fields are less often present in massive binaries than in single massive stars \citep{neiner2013}. However, the physical
details of fossil magnetic fields are only just beginning to be addressed 
\citep[e.g.][]{braithwaite2006,auriere2007,duez2010,alecian2012a,alecian2012b}.
In addition, the study of fossil fields in massive stars provides clues on the
fossil field probably also present inside cooler stars but not visible at their
surface. Therefore massive stars are a tool for the study of momentum transport
in the whole Hertzsprung-Russell diagram. 

Hot stars also represent unique targets for the study of stellar magnetospheres.
Their strong, radiatively-driven winds couple to magnetic fields and generate 
magnetospheric clouds and disks \citep[e.g.][]{babel1997,sundqvist2012}. Models
and simulations \citep[e.g.][]{uddoula2006,townsend2005,townsend2007,uddoula2013}
show that magnetic confinement of stellar winds can explain UV and X-ray
variability in magnetic OB stars \citep[e.g.][]{favata2009,petit2013}. The interaction of
the wind with the magnetic field modifies mass loss, and may lead to rapid
stellar spindown via magnetic braking
\citep[e.g.][]{weber1967,uddoula2009,townsend2010,meynet2011}. Since the
evolution of massive stars is particularly sensitive to rotation and mass loss
\citep[e.g.][]{chiosi1986,maeder2000}, the presence of a  magnetic field can
drastically influence the evolution of massive stars and thus also their
supernova explosions and feedback to the interstellar medium  \citep[ISM, see
e.g.][]{ekstrom2008}.

In addition to being structured on large scales by processes like rotation or 
magnetism, the powerful winds of hot stars can be structured on small-scales by
the intrinsic ``line-driven instability'' \citep[LDI, see e.g.][]{owocki2011}. The presence and interactions
between density structures on both these scales is poorly understood, and may
compromise the reliability of measurements of the properties of the outflows.
Moreover, most models of such line-driven winds still assume a smooth and
steady-state outflow.  Spectral diagnostics such as UV resonance and visible
recombination lines have different dependencies on density, and will provide
crucial constraints for the further development of dynamical hot star wind
models, as well as for how the resulting wind structures affect derived
quantities such as mass loss and rotation, which are essential inputs for
corresponding models of stellar evolution and feedback. Although clumping
appears to be a universal feature of line-driven winds, it is not known how the
LDI interacts with other processes that structure the wind.  Some key questions
concern possible inhibition of the lateral fragmentation of clumps, the effects
on the structure within the closed field loops, and how these different
behaviours alter the interpretation of spectral diagnostics, in particular the
determination of mass-loss rates. 

UVMag can address all these issues by providing extended time-series and
polarimetric information at higher spectroscopic resolution for OB stars,
including those already known to be magnetic. 

\subsection{Solar-type stars}\label{solar}

According to dynamo models, the variable magnetic field of the Sun is the
consequence of the interplay between two main ingredients. First the radial and
latitudinal differential rotation generate a large-scale toroidal magnetic field
from an initial poloidal field. Then the poloidal magnetic component is
regenerated. How this second process occurs is still debated, with models
invoking either the cyclonic convection in the convection zone or the transport
of decaying active regions by meridional circulation. These two steps produce a
dynamo and succeed at building continuously a large-scale magnetic field that
oscillates with time, giving rise to the 22-year solar cycle. Despite
considerable progress since the very first solar dynamo models
\citep{brun2004,charbonneau2005,brun2011}, there are still many aspects of solar magnetism that
the current models cannot reproduce or did not fully explore.

Our understanding of the solar dynamo can benefit from the observation of other
cool stars, where different dynamo types can be observed, either because they
are analogues of the Sun observed in a rare activity state (similar, e.g., to
the solar Maunder minimum) or because their physical properties (e.g. their mass
and rotation rate) differ significantly from the Sun and lead to a different
dynamo. Using spectropolarimetric observations, the magnetic fields of cool
stars can be directly characterised from the polarised signatures they produce
in spectral lines, and the associated field geometries can be reconstructed
using tomographic imaging techniques, like Zeeman-Doppler-Imaging. This has 
already been done from the ground for a small number of cool stars and provided
important results in the last decade. However, statistics are needed to obtain a
global view of the problem, i.e. we need to reach fainter stars and compare
stars in clusters with various parameters (metallicity, age...). A space-based
spectropolarimeter would allow to reach this goal thanks to a much better S/N
ratio. Associated to UV spectropolarimetry these observations would allow to
fully characterise magnetospheres around Sun-like stars, young Suns, and cool
stars in general.

The solar cycle of 22 years can be probed in other stars by observing the 
changes in activity in solar-type stars. For example the star $\tau$\,Boo has a
cycle of $\sim$2 years while the cycles detected in X-rays in HD\,81809 and
61\,Cyg\,A seem to be of the order of $\sim$8 years. An increasing number of
observational evidences for short cycles has been accumulated recently for cool
stars \citep[e.g.][]{morgenthaler2011,metcalfe2012}. Therefore a mission of 5
years or more would be sufficient to measure magnetic cycles in such stars. In
the same way as X-ray observations provided new insights on stellar cycles, UV
observations of such cycles will cast new light, in particular by allowing to
observe changes in the wind and chromosphere along the cycle. Irregularities in
the 22-year solar cycle correlate with dramatic changes in the Earth's climate
(e.g., the Maunder solar minimum probably led to the Little Ice Age on Earth),
so it could be vital to understand such long-term variations.

As the winds of cool main sequence stars are relatively weak (e.g. the solar
wind has a mass loss of $\dot{M} = 10^{-14} M_\odot$) direct detections are not
possible. Indirect detections of stellar winds in the UV can be made through
their interaction with the surrounding interstellar medium, such as the presence
of extra H{\sc i} Ly$\alpha$ absorption \citep{wood2005}. In the solar
system the heliosphere is populated by hot hydrogen atoms through charge
exchange between the ionised gas in the solar wind and the cold ISM hydrogen.
Hot hydrogen builds up particularly in the region between the bow shock and the
heliopause. In cool stars this is detected as blue-shifted Ly$\alpha$
absorption; column density and velocity measurements of this extra absorption
are fitted using hydrodynamic models of the interaction between the stellar wind
and the ISM and enable mass loss rates to be computed. Stronger winds result in
a larger astrosphere and increased absorption. Only ten systems have been
studied in this way, with new HST observations of more systems underway. With
coordinated spectropolarimetric studies it is possible to learn how the activity
levels of young solar-type stars affect the sizes of astrospheres and therefore
the strength of the winds in young planetary systems. Following the astrospheres
and magnetic fields of the stars over timescales of years will reveal how winds
and conditions inside astrospheres (i.e., in  interplanetary environments) can
change over the timeframe of stellar activity cycles.

Finally, mapping (Doppler imaging) the sizes and structures of sub-coronal
plasma in active stars requires to obtain enough signal in a few minutes so that
there is no significant smearing over the rotation period. It is also
particularly crucial to obtain observations covering at least two rotation periods
in order to discriminate between rotational modulation of stable coronal
structures and intrinsic variability (e.g., due to flares). This requires very
good phase coverage and can be done thanks to UVMag. 

\subsection{M dwarfs}

Main-sequence stars below approximately 0.35 M$\odot$ are fully convective and therefore do
not possess a tachocline, the thin shear layer at the base of the solar
convection zone thought to play an important role in generating the solar
magnetic field \citep[e.g.][]{charbonneau1997,browning2006}. Dynamo processes in
these fully-convective M dwarfs are therefore believed to differ significantly
from those in the Sun; in particular, they may operate throughout the whole
stellar interior \citep[e.g.][]{chabrier2006,browning2008}. M dwarfs are
thus of prime interest to study stellar dynamos operating in physical
conditions quite remote from the solar case as well as to understand the role of
the tachocline in the dynamos of solar-type stars.

Therefore with very cool stars, one can study both sides of the full-convection
threshold (at spectral type M4). Recent spectropolarimetric studies have shown
that partly-convective M dwarfs as well as a few fully-convective ones feature
complex magnetic geometries with a significant non-axisymmetric component
\citep{donati2008,morin2010}, while most fully-convective stars host a strong
and long-lived axial dipole component \citep{morin2008}. Explaining such a
diversity in the magnetic field geometries of M dwarfs constitutes an important
challenge for stellar dynamo theories.

Despite many differences between planetary and stellar interiors, several recent
studies have strengthened the idea that the dipole-dominated large-scale
magnetic fields observed on a number of fully-convective stars are much more
akin to planetary dynamos than to dynamos of Sun-like stars
\citep{goudard2008,christensen2009}. The discovery of the co-existence of two
distinct types of magnetism among stars having similar masses and rotation rates
among very-low-mass stars \citep{morin2010} is now interpreted in the framework
of dynamo bi-stability originally developed in the planetary dynamo context
\citep{morin2011,gastine2013}.

In addition, important evolutions of the surface magnetic fields of early M
dwarfs as well as late M dwarfs exhibiting complex fields have been observed on
timescales ranging from weeks (due to differential rotation) to years
\citep{donati2008,morin2010}. However, up to now no magnetic cycle could be
identified on these objects.

Finally, the evolution of angular momentum in M dwarfs is an issue that triggers
a number of questions. For example, \cite{reiners2012} studied the influence of
radius change across the fully-convective limit on angular momentum loss and
\cite{vidotto2011,vidotto2013} produced MHD simulations of the winds of a few M
dwarfs. Measuring wind properties of M dwarfs, e.g., astrospheric absorption
measurements in the UV (see Sect.~\ref{solar}), will provide crucial
observational constraints for these studies.

By adding spectropolarimetric capabilities to the UV spectroscopy, it will also be
possible to study flares on M dwarfs, probing the short-term variability of the
activity level of their chromospheres and the relation with the magnetic
topology.  On the most active M dwarf stars (UV Ceti type variables), flux
increases of several magnitudes in the blue/near-UV are observed on timescales
ranging from minutes to hours, and $\sim$0.1\% of their bolometric luminosity is
emitted in the form of flares. These frequent, powerful events, are caused by
magnetic reconnection. They exhibit an equivalent black-body temperature of some
10000 K and result in a number of emission lines. In particular the  hydrogen
Balmer series and Ca{\sc ii} H and K as well as lines of He{\sc i} and He{\sc
ii}, and atomic species such as Ca{\sc i}, Fe{\sc i} and Fe{\sc ii} are observed
at visible wavelengths. The UV domain  reveals a rich spectrum of highly ionised
emission lines such as C{\sc iv}, Si{\sc iv} and N{\sc v}. The most powerful M
dwarf flares can also release important amounts of high energy radiations up to
the hard X-ray spectral domain \citep{osten2010}. The radiation and particle
fluxes from flares may exert a significant influence on the atmospheres of
orbiting planets, and affect their habitability. Understanding these effects is
of prime interest for the surveys dedicated to the search for Earth-like planets
orbiting M dwarfs.

\subsection{Evolved stars}

\subsubsection{Cool supergiants}

Since both global and small-scale dynamos may be simultaneously active in the
Sun, it is not easy to disentangle the respective magnetic outcome of these two
different processes. A promising way to reach this goal consists in observing a
star with no rotation at all, or at least a star rotating so slowly that the
onset of a global dynamo in its internal layers is unlikely. If stellar
spectropolarimetry is our best asset to detect a magnetic field, the
polarimetric detection of Zeeman signatures is mostly insensitive to small-scale
magnetic elements as those expected to be generated by a local dynamo. This
issue is inescapable for solar-type dwarfs, on which millions of photospheric
convective cells are visible at any time, resulting in a highly tangled
intra-network field pattern. Cool supergiant stars may offer a rare opportunity
to circumvent this problem, since their convective cells are expected to be much
larger than on the Sun, with only a few of them covering the stellar surface
\citep{schwarzschild1975,chiavassa2010}, so that the spatial scale of
convection may be sufficiently large to limit the mutual cancellation of Zeeman
signatures of close-by magnetic elements with opposite polarities.

The feasibility of magnetic field detection in cool supergiant stars has been
successfully tested from the ground on Betelgeuse \citep{auriere2010}. Using
spatially-resolved, high-resolution UV spectroscopy with the HST,
\cite{uitenbroek1998} were able to propose a rotation period of about 17 years
and a low inclination of the rotation axis, of about 20$^\circ$. Thanks to the
brightness of  Betelgeuse, it was indeed possible to reveal the presence of a
weak $\sim$1 G surface magnetic field. This is an important observational result, in the
sense that the physical interpretations proposed for other objects to account
for their magnetic nature cannot be applied here. First, the magnetic field of
Betelgeuse has to be generated without the help of a fast, or even moderate
stellar rotation, and this specificity should exclude any global dynamo. Second,
the very large radius implies that any magnetic remnant of a strong magnetic
field on the main sequence would be too diluted to be detectable at photospheric
level. In this situation, a more natural interpretation would involve the
convection alone as the engine of a dynamo, bringing the first strong
observational evidence that such a process (initially proposed by
\cite{durney1993} and then by \cite{dorch2003} for supergiants) can be efficient
in cool stars.

This exciting result, confirmed for a larger sample of cool very bright
supergiants \citep{grunhut2010}, comes together with a number of additional
tracers of magnetic activity and convection (chromospheric emission, radial
velocities, line bisectors, Stokes V asymmetries). This wealth of information is
a motivation to pursue the spectropolarimetric monitoring of cool supergiants,
in order to investigate longer-term trends that may affect the various
measurements at our disposal and study the possible role of the surface magnetic
field in the onset of the mass-loss of Betelgeuse and other supergiant stars.
However, to study more (fainter) cool supergiants, a space-based instrument such
as UVMag is needed.

\subsubsection{AGB and post-AGB stars}

The stars located along the Asymptotic Giant Branch (AGB) are the evolutionary
descendants of low or intermediate mass stars before their transition towards
the post-AGB and the Planetary Nebulae (PN) stages.  Evolved stars located at
the tip of the AGB or present in the post-AGB domain undergo an important mass
loss which is driven mainly by radiation pressure on dust, with the supposed
combined action of other factors (e.g., condensation and opacity of dust grains,
and - for the pulsating AGB stars, namely Mira stars - stellar pulsations and
propagation of shock wave throughout the atmosphere). Hence AGB and post-AGB
objects are surrounded by rich circumstellar envelopes (CSE) exhibiting peculiar
morphologies gaining an even more and more important degree of complexity along
the rapid transition from AGB to PN symmetries \citep[e.g.][]{sahai1998}.
Magnetic fields have been invoked in order to rule the mass loss geometry and to
help to shape PN's morphology \citep{blackman2009} according to theoretical
predictions \citep{soker2002}. Throughout the last decade, several works, 
mainly based on radio-astronomy facilities, have brought observational evidences
for magnetic fields around PN and in the CSE of their AGB and post-AGB
progenitors \citep[see][for a global overview]{vlemmings2011}.

Moreover, with spectropolarimetric instruments, very weak magnetic fields (i.e.,
at the gauss level) have been detected at the surface of non pulsating M-type
AGB stars \citep{konstantinova2010} and even at the surface of a pulsating Mira
star \citep{lebre2014}. The origin of the surface magnetism in all these cool
and evolved stars (AGB/Miras, post-AGB and in red supergiants, their massive
counterparts) still has to be identified. It may rely, at least partly, within a
connection with the photospheric or the atmospheric dynamics: either by
generating a magnetic field from a local dynamo (or from the shock wave
structure itself), or by amplifying (under the action of the shock wave) an
extremely weak surface field. Within these objects, atmospheric dynamics is
indeed very complex as shown especially by high resolution spectroscopic and
interferometric observations performed for pulsating objects at very specific
dates along their light curves. Hence, while ground-based observations are
highly conditioned and limited by the weather, space facilities bring a very good
opportunity to tackle the origin of the magnetism in these evolved stars, from
AGB to PN stages.

In addition, concerning more specifically post-AGB stars, a UV facility appears
to be a powerful tool. Indeed, in globular clusters, the so-called UVbright
objects are the analogues of the field B-type post-AGB stars
\citep{moehler2001}. As they result from low-mass star evolution on their way to
the PN stage \citep{vanwinckel2003}, these objects can be confused with
post-Main Sequence hot and massive B stars. For field or cluster stars, only
detailed chemical studies (based on diagnostics from the UV and blue parts of
the spectrum) can help to identify genuine post-AGB objects, and thus to address
the problem of their magnetism.

\subsection{Additional science: ISM, novae, exoplanets}

Below we present a selection of science topics, outside of stellar physics, that
would benefit from an M-size mission such as UVMag. Would the UV and visible
spectropolarimeter of UVMag be installed on a larger telescope, other science
domains could also be considered such as Active Galactic Nuclei (AGN).

\subsubsection{ISM}

The structure and physical properties of the diffuse interstellar medium (ISM)
are best studied with high-resolution UV spectra because the resonance lines of
a wide variety of atoms and ions are located in the UV. In addition, absorption
lines formed in cold and warm ISM gas are narrow, and the velocity separation in
individual clouds along a given line of sight is often only a few km~s$^{-1}$.
Therefore, many of the observations obtained in the frame of the UVMag core
program could also be used to study the ISM. Additional observations could be
obtained in specific directions as required. Reciprocally, the study of the ISM
in the direction of UVMag's targets would allow us to better take the
correction for foreground polarization into account.

UVMag targets are best suited to study the local ISM (LISM), in the immediate
vicinity of the Sun. The LISM is composed principally of one main diffuse warm
cloud, about 10 parsecs in size, embedded in a cavity called the Local Bubble,
which may contain very tenuous, very hot gas and extends 50-150 pc around the
Sun in all directions. The Sun moves at a velocity of about 25 km~s$^{-1}$
relative to the local cloud, which is responsible for the formation of the
heliosphere by interaction with the solar wind.

The study of the local ISM is interesting to understand the structure, physical
conditions and evolution of the direct environment of our Solar System, in
interaction with it. But it is also an excellent laboratory for studying the
basic physics at work in general diffuse gas. Indeed, the simplicity of the
short sight-lines in the solar vicinity provides a unique opportunity to study
individual regions, individual clouds, individual interfaces, that are usually
blended in longer sight lines. In particular the dynamics of a cloud propagating
in a hot bubble, the interaction of hot gas and cooler material, and the
mechanisms producing O{\sc vi}, can be studied thanks to the position of the Sun
in the Local Bubble.

Another aspect of the ISM that would be ideally served by UVMag observations is
the study of molecular hydrogen in diffuse gas. The far UV domain covered by
UVMag gives rise to many H2 absorption lines in the Lyman and Werner bands.
Studies of the high rotational states (J$>$2) are particularly interesting since
there is evidence that in diffuse molecular gas these states could be
collisionally excited \citep{gry2002} and would imply the existence of a warm
phase, also related to the formation of the CH+ molecule. UVMag also gives
access to CO electronic bands present in the UV domain, and permits their
observation together with H2, for the first time at high resolution for both
species, giving the opportunity to measure the abundance ratio of these two
molecules in individual clouds and study its evolution with cloud properties.

\subsubsection{Novae}

Novae are stars that expand, and thus brighten, suddenly and for a short period
of time, due to their interaction with a close companion. They represent unique
objects to understand physical conditions of accreting matter from a companion, 
outburst, and interaction of ejecta in the ISM.

In spite of recent observational progress, two fundamental questions remain
about novae: what drives the mass loss during the outburst, and what are the
masses and structures of the ejecta? Radiative processes depend on the chemical
abundances (i.e. on evolution) and on the luminosity. These processes might
also produce a stellar wind during the ejection \citep[e.g.][]{hauschildt1994}.
Explosions are powered by the decay of radioactive elements generated after the
envelope expansion during the thermonuclear runaway, but subsequent mass loss
arises from flux distribution versus envelope opacities \citep{shore2002}.

The UV domain is the most important spectral region for the analysis of novae.
In particular in the UV it is possible to directly probe the properties
(abundances, structure, mass) of the ejecta and determine the energetics of the
thermonuclear runaway. Indeed, in the visible or IR the photometric behaviour is
driven by flux redistribution from the central remnant white dwarf, while in the
UV we can measure the resonance lines during the first months of outburst.
Measuring polarisation in these UV lines would certainly provide new insight
into the novae phenomenon.

\subsubsection{Exoplanetary magnetic fields}

From HST near-UV light-curves of transits, \cite{fossati2010} observed that the
near-UV transit light-curve of WASP-12b shows an early ingress when compared to
its transit in the visible domain. While the time of the transit egress
occurs almost simultaneously at the near-UV and visible wavelengths, the
ingress of the transit is first seen in the near-UV wavelength range. This
asymmetric behaviour has been explained by the presence of asymmetries in the
planetary atmosphere.

Close-in giant gas planets are rather inflated and most have developed an
exosphere that can fill or even overflow the planet's Roche lobe
\citep{2003ApJ...588..509G,2010Natur.463.1054L,2010ApJ...713..751I}. This may
result in mass transfer through a Lagrangian point to the star that could cause
an asymmetry in the appearance of the transiting planet-star system as seen from
the Earth \citep{2010ApJ...721..923L}. Asymmetries could also be produced by
cometary tails. However, \citet{2008A&A...483..933E} demonstrated for
HD\,209458b that a radiation-driven cometary tail would produce a late egress of
the planetary transit light curve, instead of an early ingress. In the case of
the near-UV transit asymmetry of WASP-12b, \citet{2010ApJ...722L.168V} suggested
that this asymmetry can be explained by the presence of a shock surrounding the
planet's magnetosphere. 

The interaction of a planet with the corona of its host star can give rise to
the formation of shocks that surround the planet's magnetosphere. Similar to
what occurs around the Earth and other planets in the solar system, bow-shocks
may develop around exoplanets. This idea has recently been applied to
explain the light-curve asymmetry observed in the near-UV transit of the
close-in giant planet WASP-12b \citep{2010ApJ...722L.168V}. Monte Carlo
radiation transfer simulations of the near-UV transit of WASP-12b support this
hypothesis, as it explains both the observed level of absorption and the time of
the (early) ingress observed in the near-UV light-curve of the planet
\citep{2011MNRAS.416L..41L}. 

\citet{2011MNRAS.411L..46V} applied the shock model initially developed for
WASP-12b to other known transiting systems, determining which planets are prone
to develop shocks, which could lead to an observable early near-UV ingress. They
predicted that a significant number of transiting systems (36 out of 92 planets)
might have a detectable shock, implying that bow shocks might indeed be a common
feature surrounding transiting planets. Furthermore, once the stand-off distance
of the shock (determined through the time difference of transit observations in
the near-UV and visible) and the stellar magnetic field strength are known, the
planetary magnetic field intensity can be derived.

In addition, the planetary magnetic field is believed to be responsible
for shielding the planet against the erosion of the planetary atmosphere by the
host star's wind or the impact of energetic cosmic particles. Such effects could
harm creation and development of life on the planet. Furthermore, the
presence of a planetary magnetic field may induce other sorts of
interactions, such as through reconnection between stellar and planetary
magnetic field lines. Such an interaction is believed to generate planetary
radio emission
\citep{2007P&SS...55..598Z,2010MNRAS.402.2609I,2012MNRAS.423.3285V}.
Unfortunately, despite many attempts, radio emission from exoplanets has not
been detected so far. 

Near-UV observations during planetary transits may provide an alternative to
probe planetary magnetic fields over observations at radio wavelengths. However,
the observations of these sorts of systems requires regular observations to
monitor the activity level. Coordinated observations of systems in the UV and
visible domains would be important to study the environment and fate of
hot Jupiters. Therefore, clear synergies exist between UVMag and targets
observed with space missions dedicated to exoplanets such as CHEOPS, TESS
or Plato.

\section{The UVMag space project}

\subsection{UV and visible spectropolarimetry}

To fulfil the requirements of the scientific objectives presented above, we
propose to develop a UV and visible spectropolarimeter on an M-size space
mission, with a telescope diameter of about 1.3 meters. Since many of the
targeted phenomena are known, or at least suspected, to be sensitive to
metallicity, it is essential that stars in the LMC and SMC can be
reached. This objective is just fulfilled with the proposed baseline aperture
of 1.3 m, which should, therefore, not be reduced.

The spectropolarimeter should ideally cover the full wavelength range from 90 to
1000 nm and at least the most important lines in the domains 117-320 nm and
390-870 nm. Polarisation should be measured at least in Stokes V (circular
polarisation) in spectral lines, but the aim is to measure all Stokes QUV
parameters (circular and linear polarisation) in the lines and continuum. A high
spectral resolution is required, at least 25000 in the UV domain and at least
35000 in the visible, with a goal of 80000 to 100000 especially in
the far-UV (90-117 nm) to increase the spatial resolution and improve the
ISM studies. The peak signal-to-noise ratio should typically be above 100.

Spectroscopy with these specifications in the UV and visible domains is
relatively easy to achieve with today's technology and detectors. A 
preliminary design of the UVMag echelle spectrograph for the wavelength range
117 to 870 nm has already been done, using a grism for the UV domain and a prism
for the visible domain. In the visible domain, the detector will be a thinned
back-illuminated double-depletion CCD passively cooled to below -60 degrees. In
the UV domain, two different types of detectors will probably have to be used: a
back-illuminated CCD for the longer UV wavelength range and multi-channel plates
(MCP) with a multianode readout device for the shorter UV wavelength range.

However, spectropolarimetry (rather than just spectroscopy) in the UV and
visible domains is more challenging. Indeed, (1) high-resolution
spectropolarimetry of stars has never been obtained from space; (2) visible spectropolarimeters available on the ground are large; and (3) it is
very important to keep the instrumental polarisation at a low level (below a few
percent). In addition, several technical issues need to be addressed, such
as the lack of birefringent material over the full UV domain, the wish to avoid
moving parts in space, and the difficulty to assemble optical components while
keeping them transparent to UV light (i.e. without using glue). Therefore we
have started a Research \& Technology (R\&T) program to study a space UV+visible spectropolarimeter. Our study is based on existing ground-based
spectropolarimeters, such as ESPaDOnS or Narval, and new spectropolarimetric
techniques proposed in the literature. In particular, we are studying two
possible designs for the polarimeter, one combining spatial and spectral
modulation following  \cite{sparks2012} and the other one using polychromatic
temporal modulation adapted from \cite{snik2012}. See \cite{pertenais2014} for
more details on the polarimeter study and its challenges.

Information about the UVMag design can be found at http://lesia.obspm.fr/UVMag
as the project progresses.

\subsection{Observing program}

UVMag will observe all types of stars in the magnitude range at least V=3-10.
The observing program includes two parts: (1) 50 to 100 stars will be
observed over at least one full rotational cycle with high cadence in
order to study them in great details and reconstruct 3D maps of their surface
and environment. In addition, the solar-like stars among those will be
re-observed every year to study their variability over activity cycle
timescales. We propose that the stars will be selected following a
proposal process, the data will be distributed to the corresponding teams 
and become public after a 1-year proprietary period; and (2) one or two
spectropolarimetric measurements of several thousands stars will be
obtained to provide information on their magnetic field, wind and environment.
This will include an unbiased statistical survey as well as targets
selected following a proposal process. The unbiased survey data (intensity and
Stokes spectra) will be made publicly available upon acquisition. These
snapshot data will provide input for stellar modelling.

The acquisition of the data for these programs will take 5 years. A
longer mission would of course allow us to observe more targets, in particular
to propose interesting targets detected in the survey sample for detailed
follow-up, and to expand the coverage of activity cycles.

\section{Conclusions}

We plan to study how fossil magnetic fields confine the wind of
massive stars and influence wind clumping, how magnetic interactions impact
binary stars, how a solar dynamo impacts the planets and how it evolves, how
magnetic fields, winds and mass-loss influence the late stages of stellar
evolution, in which conditions a magnetic dynamo develops, how the angular
momentum of stars evolves, how small-scale and large-scale stellar dynamos work
and how their cycles influence their environment, what explains the diversity of
magnetic properties in M dwarfs, what causes the segregation of tepid stars in
two categories (those with sub-gauss magnetic fields and those with fields above
a few hundreds gauss), what are the timescales over which magnetospheric
accretion stops in PMS stars, etc. These questions will be answered by observing
all types of stars: massive stars, giants and supergiants, chemically peculiar
stars, pre-main sequence stars, cool stars, solar twins, M dwarfs, AGB and
post-AGB stars, binaries, etc. Additional possible science includes the study of
the ISM, white dwarfs, novae, exoplanets, atomic physics,...

The UVMag consortium has set the basic requirements for an M-size space mission
to study the magnetospheres and winds of all types of stars. This is the next
step to progress on the characterisation and modelling of stellar environments,
as well as on important questions regarding stellar formation, structure and
evolution. Simultaneous UV and visible spectropolarimetry over long
periods of time is indeed the only way to comprehend the full interaction
between various physical processes such as the stellar magnetic field and
stellar wind. An R\&T  study is ongoing for the instrument. The M-size
mission will be proposed at ESA. We also consider the option of installing the
spectropolarimeter of UVMag as part of a series of instruments on a large UV and
visible observatory, such as NASA's LUVOIR project (see NASA's
astrophysics
roadmap\footnote{http://science.nasa.gov/media/medialibrary/2013/12/20/secure-Astrophysics\_Roadmap\_2013.pdf}).

\acknowledgments
The UVMag R\&T program is funded by the French space agency CNES.

\bibliographystyle{spr-mp-nameyear-cnd}  
\bibliography{articles}                

\end{document}